\documentclass[aps]{revtex4}
\usepackage{amssymb,amsmath,bm,enumerate,graphicx,color,subfigure}
\newcommand{\be}{\begin{equation}}
\newcommand{\ee}{\end{equation}}
\begin{document}
\title{ Path-wise versus  kinetic  modeling  for  equilibrating   non-Langevin  jump-type processes}
\author{Mariusz  \.{Z}aba,  Piotr Garbaczewski  and  Vladimir Stephanovich}
\affiliation{Institute of Physics, University of Opole, 45-052 Opole, Poland}
\date{\today }
\begin{abstract}
We discuss  two independent methods of solution of a master equation whose biased
 jump transition rates account for long jumps  of   L\'{e}vy-stable type and
  admit a Boltzmannian  (thermal)  equilibrium to arise in
   the large time asymptotics of a probability density function  $\rho (x,t)$.
 Our  main  goal   is to  demonstrate  a  compatibility of    a
 {\it direct} solution  method   (an explicit, albeit  numerically assisted,
  integration of the master equation) with  an  {\it indirect}  path-wise  procedure,  recently
proposed   in  [Physica {\bf A 392}, 3485, (2013)] as a valid tool
for a dynamical analysis of non-Langevin jump-type processes. The
path-wise method heavily relies on an accumulation of large sample
path data, that are generated by means of a properly tailored
Gillespie's algorithm. Their statistical analysis in turn allows  to
infer  the dynamics of $\rho (x,t)$.  However,  no consistency check
has been completed so far  to demonstrate that both methods are
fully compatible  and indeed provide a solution of the same
dynamical problem. Presently we remove  this gap, with a focus on
potential deficiencies (various cutoffs, including those upon the
jump size) of approximations involved in simulation routines and
solutions protocols.
 \end{abstract}
\maketitle

\section{Introduction.}

Investigating  random dynamics is  nowadays  an interesting
interdisciplinary research  field,   particularly because  the same
patterns of dynamical behavior can be  identified and  traced over
in different random systems.  Here the term "different" pertains
both  to scientific disciplines where an impact of randomness is to
be  analyzed and to various facets of  randomness itself (diffusion
versus jump process issue  is  here a drastic oversimplification).
Concerning disciplines involved, those range from economics
\cite{stanley} and theory of networks \cite{dor} to biology  and
physics \cite{dor,vank, levy1}.

Quite varied   randomness manifestations  were either explicitly
identified  or theoretically and numerically tested   in a a  large
number of physical systems.   Among them we indicate glassy and
porous media,  nano-systems like e.g. nano-ferroics (nano-structured
substances with magnetic, electric and/or combined long-range
orders, \cite{book}). There,  stochastic features  need to be
accounted for  on the mesoscopic nanometer-sized level of basic
(coarse-grained) constituents. The randomness (noise) input cannot
be neglected if  observable  physical properties of the system are
quantified and next confronted with experimental data.

In random systems,  standard analytical methods based on presuming
detailed microscopic  models of stochastic dynamics  (typically
Langevin-based), \cite{mezard}, arise as an outcome of a number of
simplifying  steps. Therefore a potentially  complete information
about the system  is lost, as being  severely limited by the
condition of analytical tractability.

 If we wish to relax the
tractability assumption, then  to get reliable results one typically
needs to turn over to  direct numerical simulations (like the
Monte-Carlo method, \cite{mc}). An obvious advantage of a numerical
assistance is that a statistical analysis of accumulated sample
paths data allow  to test a congruence between theory and experiment
for any random system under consideration. Moreover, various  path
ensemble mean values can be confronted with the single trajectory
statistical features, like e.g.  in connection with  an ergodicity
issue.

Many problems of stochastic dynamics   can be analyzed by means of
the "mean-field"  smoothing out of  microscopic quantities  (random
variable in the Langevin equation) and passing to  macroscopic ones.
That amounts to deducing a suitable differential or
integro-differential equation for the probability density function
(pdf) $\rho({\bf r},t)$ (or $\rho(x,t)$ in one spatial dimension)
whose numerical solution is seemingly easier and typically can be
handled on a standard personal computer.

The present paper  departs from  a specific class of jump-type
noises and  stochastic  processes  whose prototype pdfs   determine
instantaneous spatial displacements (jumps) with a definite
predominance of  large ones (heavy-tailed  pads and measures
appropriate for symmetric  L\'{e}vy flights), \cite{levy1}.

In the presence of conservative  forces,   effects of  L\'{e}vy
noise and its response to external potentials are  normally
quantified in terms of the appropriate  Langevin equation,
\cite{fogedby,metzler}.
 However, it is well known (c.f. the Eliazar-Klafter no-go statement,
\cite{klafter}) that  the pertinent random motion  can  never
asymptotically  set  down at thermal equilibrium pdf of the
Boltzmann form.

On the other hand,  we have a clear physical motivation for our
present  research. A  large class of non-Langevin  L\'{e}vy-type
stochastic  processes has been identified, \cite{thermal,gar1,gar2},
that  admit thermal equilibria in the canonical (Boltzmann) form,
but theirs  sample paths cannot be deduced wthin he Langevin
picture. These processes generically refer to the random dynamics in
inhomogeneous media, as openly stated in Ref. \cite{gar1}.

 In this case,   the main problem previously   addressed in Ref. \cite{zaba1}
    was to infer  a direct   non-Langevin path-wise
description of random motion, on the sole   basis of the pre-defined
master equation.  The task has been accomplished by means of a
properly tailored Gillespie's algorithm.  Sample paths have been
 found to show up a "qualitative typicality",  properly reflecting  the
   imbalance  (e.g. potential  predominance of large  versus small
jumps), for different values  of the stability index $\mu \in (0,2)$
in the related L\'{e}vy measure.

The main question, posed (and answered) in the present work,  is
whether the pdf   $\rho (x,t)$   evolution, statistically inferred
from large sample path data  (the latter  obtained by  means of
direct numerical simulations)  is equivalent to that obtained from a
direct numerical solution of the corresponding master equation. Once
the answer is set to be positive,
  the  next  issue  of interest is:  which of those solution methods turns   out to be
  more efficient (i.e. effectively  faster).  Our  focus  is on the former
  issue, while the   latter (efficiency) topic will  be briefly
  commented.

We quantify a pdf $\rho(x,t)$ evolution, that is   driven by a non
-Langevin jump-type process on $R$,    in terms of   the master
equation, \cite{zaba1,geisel,belik}:
\begin{equation}
\partial_t\rho(x,t)=\int  [w_\phi(x|y)\rho(y,t)-w_\phi(y|x)\rho(x,t)]dy,\label{l1}
\end{equation}
where an integral is interpreted in terms of its Cauchy principal value and
\begin{eqnarray}
w_\phi(x|y)&=&C_\mu  \, \frac{\exp[(\Phi(y)-\Phi(x))/2]}{|x-y|^{1+\mu}},\nonumber \\ \nonumber \\
C_\mu&=&\frac{\Gamma(1+\mu)\sin(\pi\mu/2)}{\pi} \label{l2}
\end{eqnarray}
is the  jump   transition rate  from  $y$  to  $x$. We assume to
have given a priori a "potential landscape" represented by a
(Newtonian potential)   function $\Phi (x)$, $x\in R$ and  thence an
asymptotic invariant  (Boltzmannian)   pdf $\rho _*(x) \sim \exp[-
\Phi (x)]$ of the random   process governed by Eq. (1).   A concrete
 L\'{e}vy    noise input is   identified by a  stability index    $\mu \in (0,2)$   of   the involved   L\'{e}vy measure  $ \nu _{\mu } (dx) =
 [C_{\mu }\,   \, /|x|^{\mu }]\, dx$, \cite{sato,lorinczi}.
  As  a  standard initial data choice for the pdf  dynamics we take a narrow Gaussian  (mimicking   the  Dirac  delta, e.g. a
    point-like source), $\rho (x,0)= (2\pi \sigma ^2)^{-1/2} \, \exp [-x^/2\sigma ^2]$ with $\sigma ^2 = 10^{-3}$.

We note  that  $w_\phi(x|y) $    typically    is a   non-symmetric function of $x$ and $y$.   This   needs  to be contrasted with
any    "free"  symmetric    L\,{e}vy-stable  noise (take $\Phi  =0  $ identically)     whose transition rates are  $(x,y)$- symmetric.

As   long as we are interested only in  a  solution  $\rho (x,t)$   of the master equation, we tell about a   {\it kinetic modelling}
 of the underlying stochastic process, even if its   minute   details (like  e.g. the random variable and related sample paths )
  are not  a priori defined. On the other hand,    if   one  can  devise a   path-wise picture of the
  underlying   random  dynamics,    whose  statistical    consequence     is the  evolution  (1)    of $\rho (x,t)$,  we tell about  an
{\it    indirect path-wise  method}  of solution of   the master equation.  The latter case was the subject of our recent paper \cite{zaba1},
 followed by a straightforward extension
   from 1D to 2D, \cite{zaba2}.

  Since  generically    there is no Langevin representation  of the  random  dynamics in question, our
main goal    in Ref. \cite{zaba1}   has been   to establish the
appropriate path-wise description of the underlying jump-type
process and  subsequently   to  infer the $\rho(x, t)$ dynamics
directly from the   random paths statistics  data. However, it is
not  at all obvious, albeit  anticipated in Ref. \cite{zaba1}, that
the path-wise route  gives  rise to  the same evolution of  the
probability distribution as that coming out from  a direct
integration of  the   master equation (1).  A reliable validity test
is necessary here.

To deal with a tractable dynamical problem,   both  the indirect
path-wise procedure  of Ref. \cite{zaba1}, and the direct  method
will be  resolved in terms of  a truncated   jump-type process on
$R$, c.f. \cite{mantegna}. (In passing we note that there is no
jeopardy  to end up with the Gaussian pdf in the large time
asymptotics, which is a property of standard   jump processes
\cite{mantegna}.  Our biased transition rates definitely force the
random system to set down at non-Gaussian pdfs, even with cuttoffs
upon the jump size.)

  To this  end we    impose suitable cutoffs upon the original    master equation (1),  hereby rewritten in a more handy form  ($x-y=z$) :
\begin{equation}
\partial_t\rho(x,t)=\int\limits_{\varepsilon_1\leq|z|\leq \varepsilon_2} \biggl[w_\phi(x|z+x)\rho(z+x,t)
-w_\phi(z+x|x)\rho(x,t)\biggr]dz.\label{l3}
\end{equation}
where $\varepsilon_1$ and $\varepsilon_2$ are  a priori  chosen,
respectively  lower and upper bounds   for the     jump size;
$\varepsilon_1 = 0.001$ and $\varepsilon_2 = 1$  have been adopted
in Ref. \cite{zaba1}.

 Assuming that the time increment $\Delta t$ is small enough, we  can  rewrite the  master equation
 in an approximate form
\be \rho(x,t+\Delta t)\approx\rho(x,t)+\Delta
t\int\limits_{\varepsilon_1\leqslant|z|\leqslant \varepsilon_2}
[w_\phi(x|z+x)\rho(z+x,t)-w_\phi(z+x|x)\rho(x,t)]dz.\label{l3a} \ee
whose iteration in the finite time interval $t\in [0,T]$ is expected
to give positive-valued  outcomes  $\Delta t \ll 1$ removes a
jeopardy of negative  values,  after a number of iteration steps).

The  numerically assisted iteration procedure   needs one more
cutoff. Namely, for each particular  choice of an  asymptotic pdf
$\rho _*(x)$  we  need to establish  a fine-tuned  fidelity interval
$[-a,a]$, $a>0$, within which a substantial "probability mass" is
concentrated.    Effectively, this   fine tuning  amounts to
selecting $a$  so  that $\rho _*(\pm a)$ is sufficiently close to
zero so that a contribution from the complement (exterior) of
$[-a,a]$ in $R$  can be justifiably disregarded. A concrete
parameter $a$ selection depends on the specific pdf $\rho _*(x)$
under consideration.

Our numerical calculation  requires  a partitioning of  $[-a,a]$ into
 small  pieces and storing them in a table     of $\rho(x,t)$ values
  at every mid-point of  that partition.
  The partition grid can not be too fine as in latter case
 the standard Simpson's  integration in Eq. \eqref{l3} unnecessarily
becomes exceedingly  time consuming.  The speed of simulation is
inversely correlated with the  number of encountered partition
mid-points, i.e. the partition finesse.

{\bf  Comment:}  To make the present paper self-contained, here we
briefly recapitulate the essence of our direct path-wise modelling
method of  the  pdf  dynamics  \eqref{l1}. We refer the reader to
Ref. \cite{zaba1} for details. Our numerical method relies heavily
on Gillespie's algorithm, devised to model the dynamics of chemical
reactions. In our procedure, chemical reaction channels of the
original Gillespie's algorithm have been re-interpreted as jumps
from one spatial point  to another, like transition channels in the
spatial  jump process. An obvious provision is that the set of
possible chemical reaction channels is finite (and generically low),
while we are interested in all admissible jumps from a chosen point
of origin  $x_0$ to any of
 $[x_0-\varepsilon_2,x_0-\varepsilon_1]\cup[x_0+\varepsilon_1,x_0+\varepsilon_2]$.
  It is clear  that such jumps form an infinite  continuous set.
  It is obvious that in the numerical algorithm we cannot admit all
   conceivable jump sizes. As well, the number of destination points,
    even if potentially enormous, must remain finite for any fixed point of origin.

Our modified version of the  Gillespie's algorithm, appropriate for
handling of such  spatial  jumps reads  as follows:
\begin{enumerate} [(i)]
\item Set time  $t=0$ and the point of origin  $x=x_0$.
\item Create the set of all admissible  jumps from $x_0$ to $x_0+z$ that is
compatible with the transition rate  $w_\phi(z+x_0|x_0)$. \\
\item Evaluate
\begin{eqnarray}
&&W_1(x_0)=\int_{-\varepsilon_2}^{-\varepsilon_1}w_\phi(z+x_0|x_0)dz, \nonumber \\
&&W_2(x_0)=\int_{\varepsilon_1}^{\varepsilon_2}w_\phi(z+x_0|x_0)dz
\label{l4a}
\end{eqnarray}
and  $W(x_0)=W_1(x_0)+W_2(x_0)$.
\item Using  a random number generator draw  $p\in[0,1]$  from a uniform distribution.
\item Using above $p$   and identities
\begin{widetext}
\begin{equation}
\left\{
  \begin{array}{ll}
    \int\limits_{-\varepsilon_2}^{b}w_\phi(z+x_0|x_0)dz=p W(x_0), & \hbox{$p<W_1(x_0)/W(x_0)$;} \\
    W_1(x_0)+\int\limits_{\varepsilon_1}^{b}w_\phi(z+x_0|x_0)dz=p W(x_0), & \hbox{$p\geqslant W_1(x_0)/W(x_0)$,}\label{l5a}
  \end{array}
\right.
\end{equation}
\end{widetext}
find  $b$ corresponding to the  "transition channel"  $x_0
\rightarrow  b$.
\item  Draw a new number  $q\in(0,1)$  from a uniform distribution.
\item Reset time label  $t=t+\Delta t$  where  $\Delta t=-\ln q/W(x_0)$.
\item Reset $x_0$ to a new value $x_0+b$.
\item Return  to step (ii) and repeat the procedure  anew.
\end{enumerate}

We point out that  the item  (vii) in the above is most relevant for
the  explanation of the ultimate efficiency of path-wise
simulations. The   time interval is not uniquely fixed  for the
whole numerical procedure and   directly depends on the  current
state of the system.

\section{Analysis  of kinetic and path-wise   outcomes  for various target  pdfs   and   L\'{e}vy   drivers.}

\subsection{Gaussian target (harmonic confinement)}

\begin{figure}[h]
\begin{center}
\centering
\includegraphics[width=90mm,height=90mm]{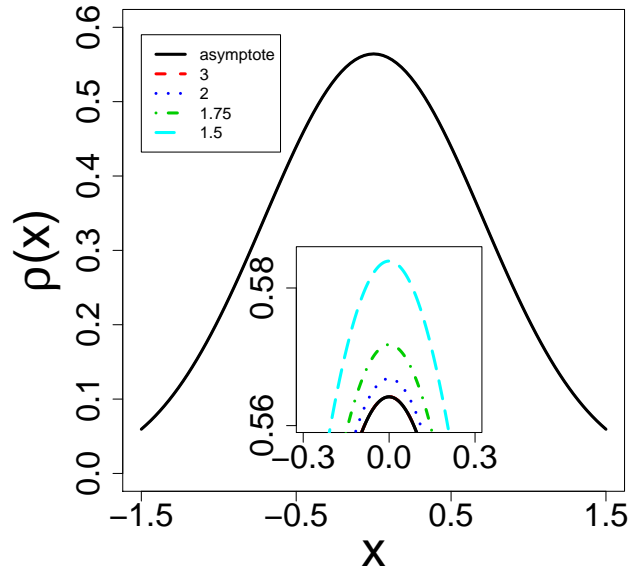}
\caption{Gaussian target:    $x$-dependence of   $\rho(x, t)$,  evaluated directly from  the  master equation (3),   for   $\mu = 1$ and  the terminal time instant  $t=15$. An
 initial condition is  $\rho(x, 0) \sim\delta(x)$,  c.f. main text.
 The inset  (note the scale change) quantifies minor  differences between  directly evaluated  $\rho _a(x,t=15)$
  for  different   $a$ values, shown in the legend.  The red line ($a=3$) in the main panel and inset cannot be distinguished from the asymptote
$\rho _*(x)$, shown in black.}
 \end{center}
\end{figure}

\begin{figure}[h]
\begin{center}
\centering
\includegraphics[width=120mm,height=120mm]{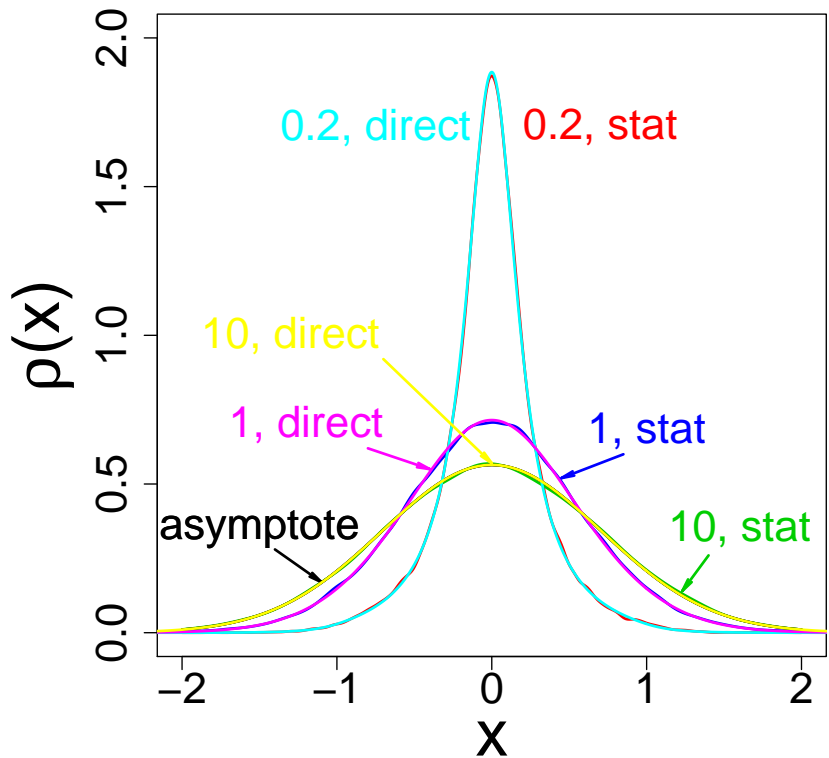}
\caption{Gaussian target: Comparison of the time evolution of $\rho(x, t)$ inferred from 75 000 trajectories (indicated by {\it stat})
 and directly from (\ref{l3}) (indicated by {\it direct}). We  reproduce $\mu = 1$  case only. The outcomes of  the direct and path-wise methods are
 almost indistinguishable in the adopted scale. All qualitative
features are
 preserved in case of  $\mu = 0.5$  and $\mu = 1.5$ (not reproduced).}
\end{center}
\end{figure}

Let us consider an asymptotic invariant (target) pdf in the Gaussian
form
\be \rho_*(x)=\frac{1}{\sqrt{\pi}}e^{-x^2}.\label{l4} \ee
 The
corresponding $\mu$-family of transition rates reads
\be
w_\phi(z+x|x)=C_\mu\frac{e^{-z^2/2-x z}}{|z|^{1+\mu}}.\label{l5}
 \ee

Fig. 1 displays a dependence  of the terminal pdf     $\rho
(x,t=15)$  upon the  choice of   $a$   (e.g. the  fidelity interval
boundary) for the  pdf $\rho _*(x)$.  We consider the Cauchy driver
$\mu=1$. With the increase of  the width of  the fidelity interval,
the pdf $\rho (x,t)$  approaches smoothly an invariant density $\rho _*(x)$ of Eq. (5).
 In case of     $a=3$  the outcome   depicted in  red,
 for all practical purposes (fapp)  is   indistinguishable from  the Gaussian, depicted in black.

\begin{figure}[h]
\begin{center}
\centering
\includegraphics[width=50mm,height=50mm]{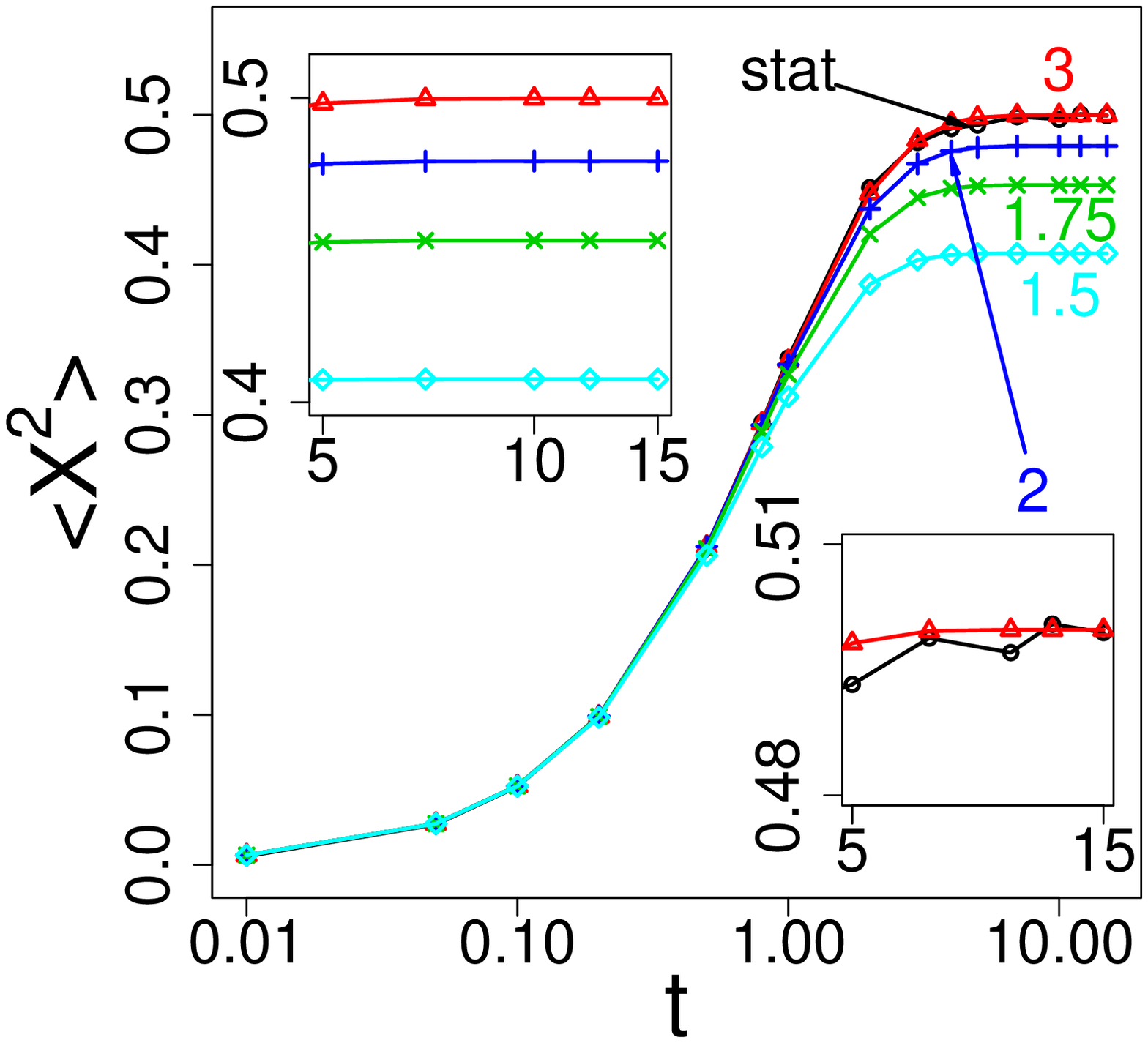}
\includegraphics[width=50mm,height=50mm]{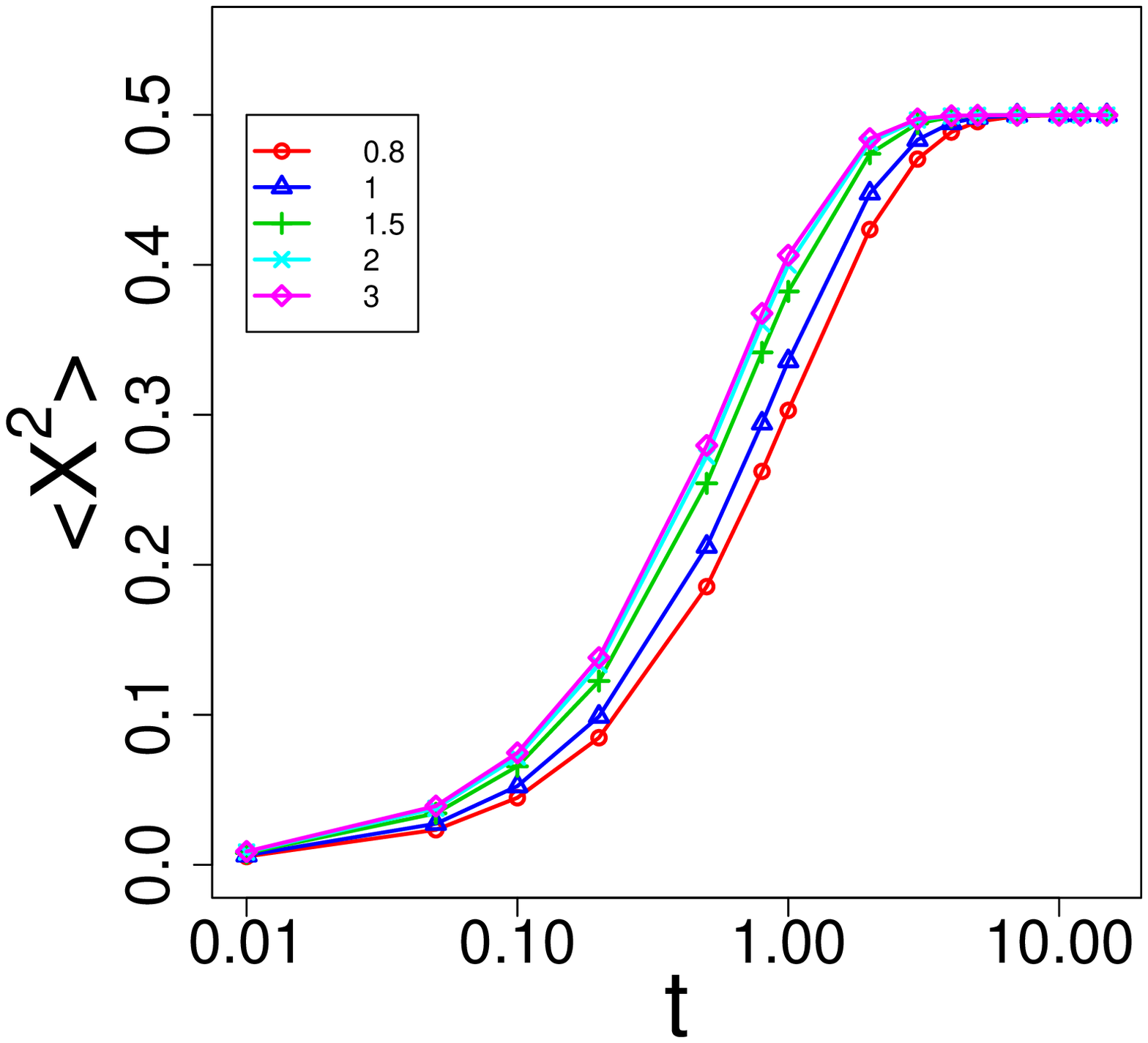}
\includegraphics[width=50mm,height=50mm]{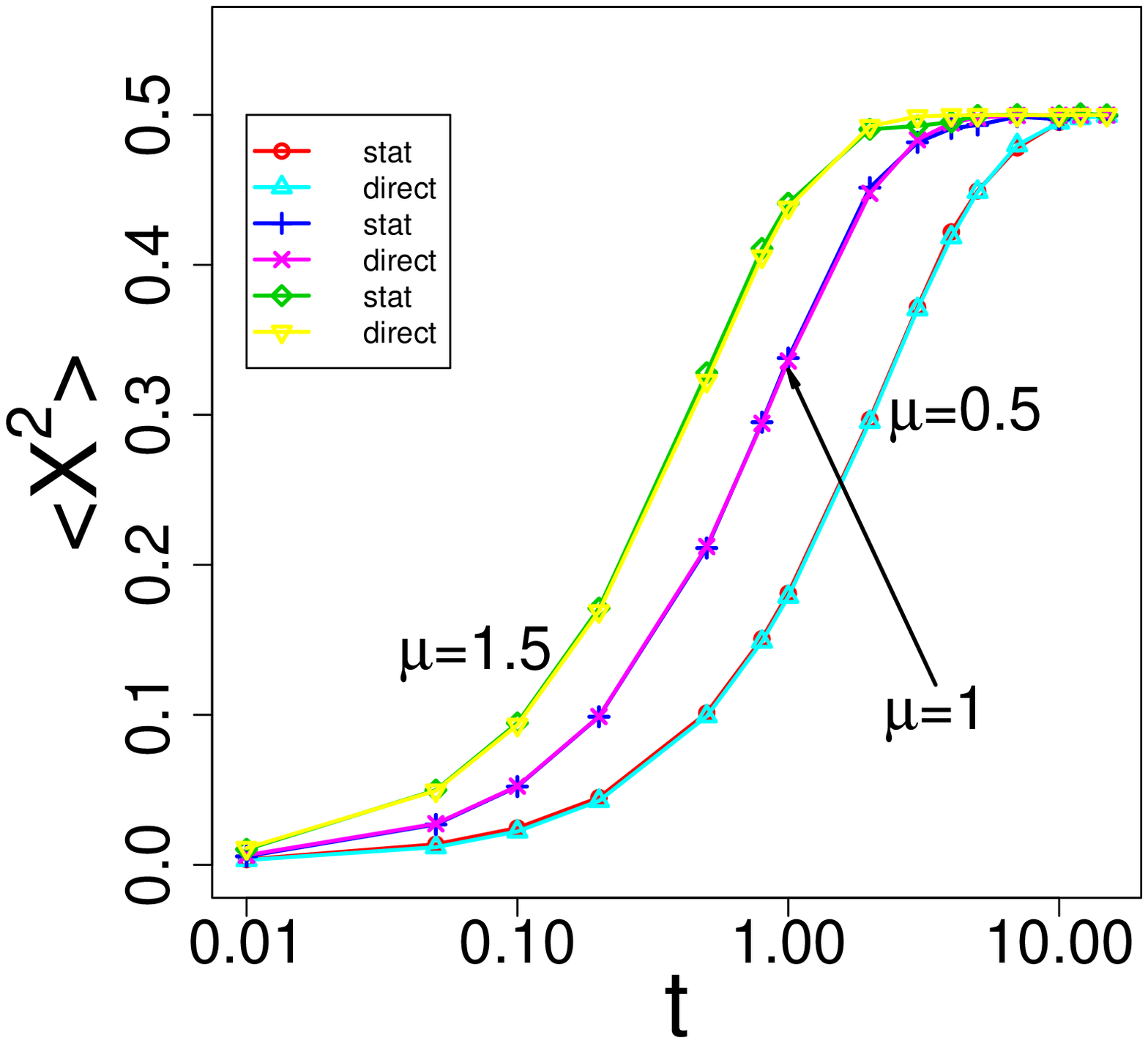}
\caption{Gaussian target: Time evolution of the pdf $\rho(x, t)$ second moment: dependence on $a$ for $\mu=1$ (left panel),
dependence on various $\varepsilon_2$ for interval $[-3,3]$ and $\mu=1$ (middle panel),
comparison of the two methods for various $\mu$ on $ [-3,3]$ (right panel).
 All trajectories and probability distribution  were started from a point-like source $ \sim \delta(x)$.}
\end{center}
\end{figure}

Fig. 2. displays a probability density evolution,   directly  computed from  (3)  and (4) for  a fixed fidelity   interval $[-3,3]$.
The outcomes are compared with those inferred via an indirect path-wise  method  from the ensemble statistics of 75000 trajectories. Both methods  clearly  give consistent results.

 Fig. 3 depicts the time dependence of the second moment   of the involved   pdf $\rho(x,t)$.  The  left panel in Fig. 3 displays a dependence  of $\rho (x,t)$  on different  fidelity interval boundary values  $a$ for $\mu=1$. That is
 compared  (path-wise method)  with the  statistical outcome provided by  $75 000$ trajectories  ({\it stat}   indication).    A numerical convergence to $< X^2 >_\infty=1/2$ is consistent with an analytic equilibrium value of the second
moment of the a priori prescribed   Gaussian  target  $\rho_*(x)$ . Note   that   the  assignment  of   too small   value
     $a>0$ modifies the convergence of the second moment:   limiting  values $<1/2$  would necessarily appear.  Indeed:
\be
<X^2>_a=\int\limits_{-a}^ax^2\rho_*(x)dx=\left\{
                             \begin{array}{ll}
                               0.5, & \hbox{$a=\infty$;} \\
                               0.49978, & \hbox{$a=3$;} \\
                               0.47699, & \hbox{$a=2$;} \\
                               0.44716, & \hbox{$a=1.75$;} \\
                               0.39385, & \hbox{$a=1.5$.}
                             \end{array}
                           \right.\label{l6}
\ee
A middle panel in Fig. 3 displays the time dependence of the pdf
$\rho(x,t)$ second moment for various $\varepsilon_2$   (upper bound upon the jump size), once the fidelity interval  $[-3,3]$ has been  adopted.
  It is clear that with the increase of   $\varepsilon_2$,   the  asymptotic
pdf (\ref{l4}) is approached  faster.

 The pertinent time  rate of  convergence towards an
asymptotic pdf cannot  be  made  too  large, unless the fidelity interval gets  much  larger.
One may  choose  $\varepsilon_2  >3$ , while   keeping  intact    $a=3$. There is no   change  in   the  second moment  temporal   behavior.
The right panel in Fig. 3  confirms that both    direct and path-wise  methods give the same evolution of   the second moment,
for three specific choices of L\'{e}vy drivers, namely $\mu =  0.5, 1, 1.5$.

\subsection{Quadratic Cauchy target (logarithmic confinement)}

\begin{figure}[h]
\begin{center}
\centering
\includegraphics[width=55mm,height=55mm]{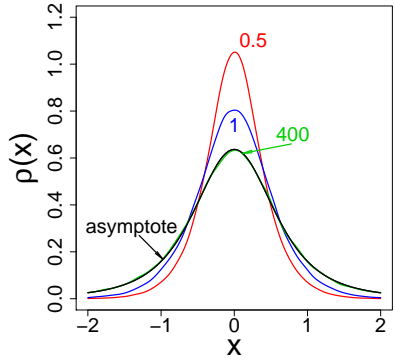}
\caption{Quadratic Cauchy target:  time evolution of $\rho(x, t)$ evaluated directly from  the  master equation (3),  for the fidelity interval  $[-20,20]$. The  $\rho(x)$    path-wise
 inference  from $200 000$ trajectories  (not reproduced, see however \cite{zaba1})  gives  curves   that cannot be distinguished from  the  former outcome.}
\end{center}
\end{figure}
\begin{figure}[h]
\begin{center}
\centering
\includegraphics[width=50mm,height=50mm]{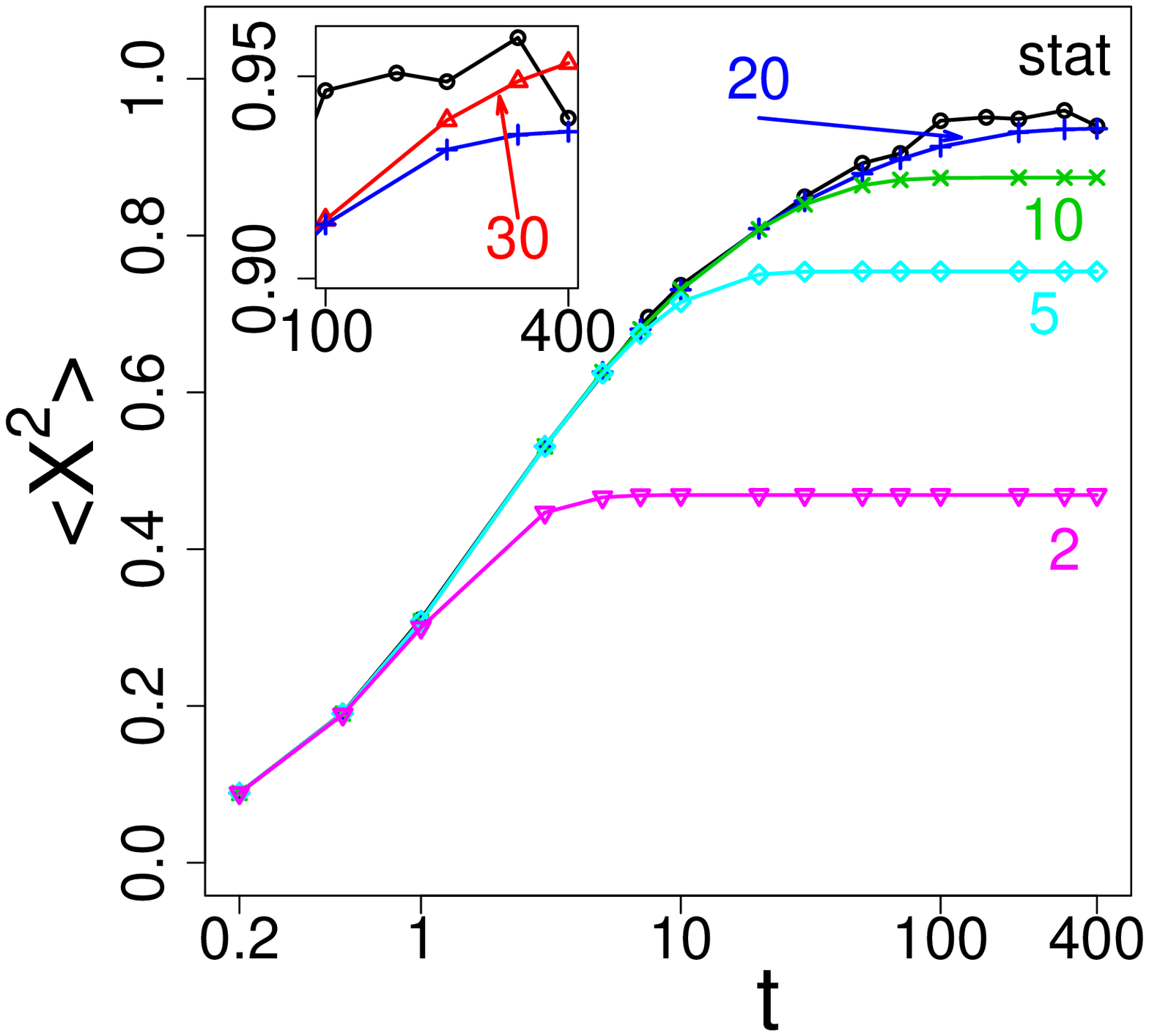}
\includegraphics[width=50mm,height=50mm]{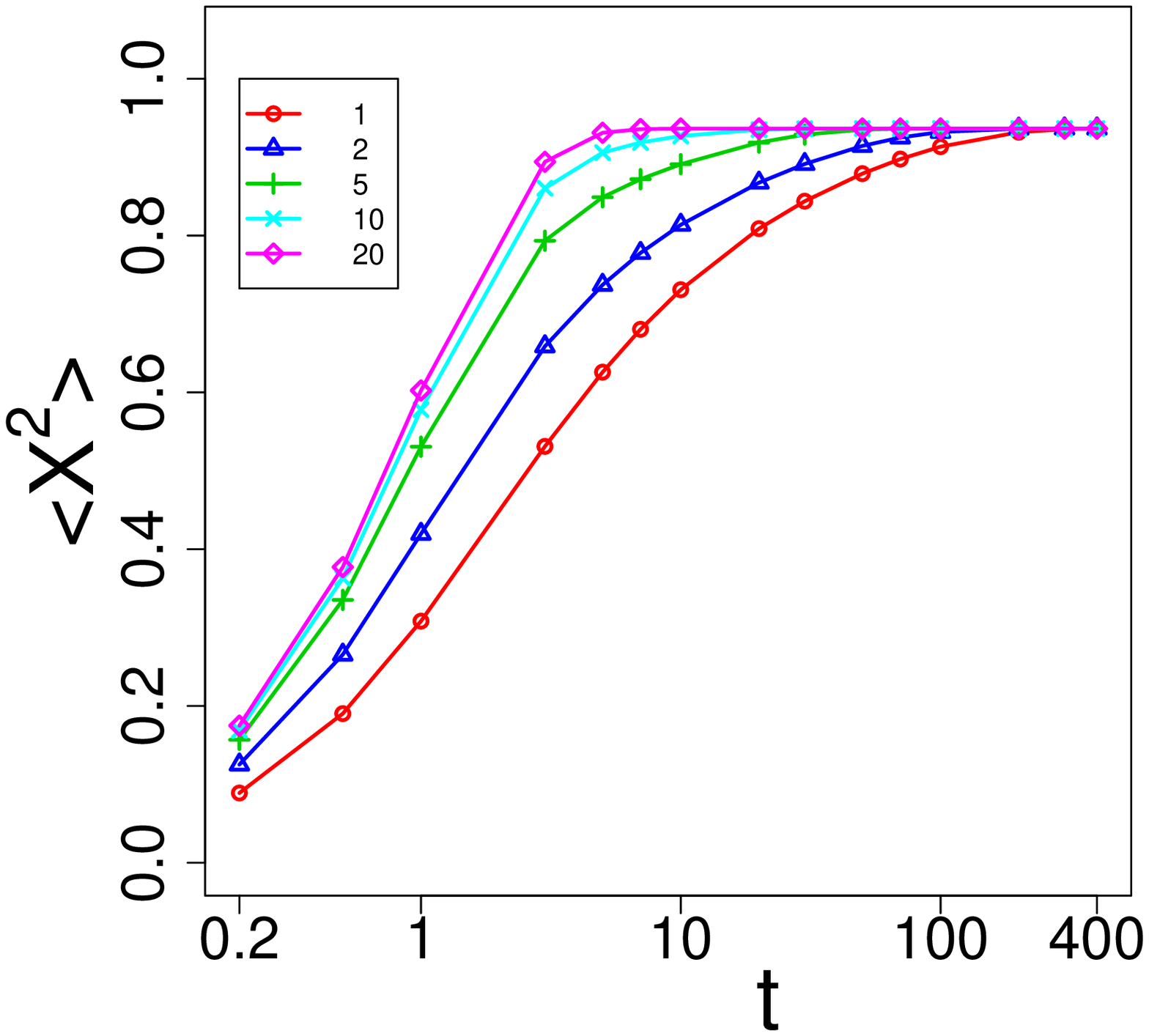}
\includegraphics[width=50mm,height=50mm]{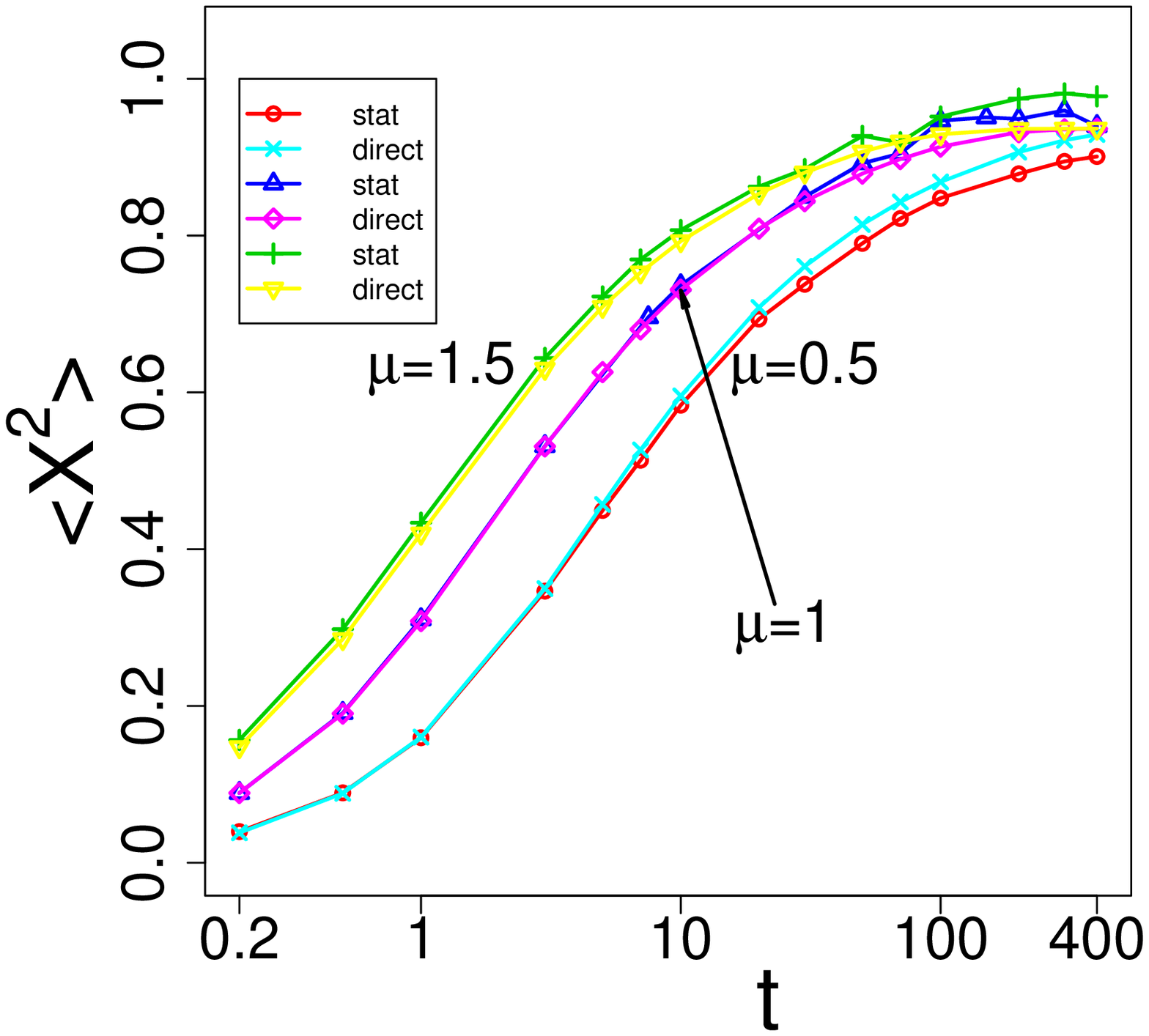}
\caption{Quadratic Cauchy target.  Time evolution of the pdf $\rho(x, t)$ second moment. Dependence on $a$ for $\mu=1$ (left panel), dependence on various $\varepsilon_2$ for  the fidelity  interval $[-20,20]$
 and $\mu=1$ (middle panel). Comparison of  two  solution   methods for various $\mu$ and  $[-20,20]$ (right panel). All trajectories and probability distribution  were started  point-wise  $\rho (x,0) \sim \delta(x)$.}
\end{center}
\end{figure}

Let us consider an asymptotic invariant (target) pdf in form
\be
\rho_*(x)=\frac{2}{\pi}\frac{1}{(1+x^2)^2}.\label{l7}
\ee
The transition rate (\ref{l2}) $w_\phi(z+x|x)$ for any $\mu\in(0,2)$
takes the form
\be
w_\phi(z+x|x)=\frac{C_\mu}{|z|^{1+\mu}}\frac{1+x^2}{1+(z+x)^2}.\label{l8}
\ee
Fig. 4 displays a pdf evolution that has been  directly  deduced from  Eq.  (3), for  the fidelity interval    $[-20,20]$. The path-wise outcome (not reproduced here, see \cite{zaba1})
from the ensemble statistics of   200 000 trajectories  gives  the pdf evolution which cannot be distinguished from the former within adopted scales.

The second moment of the present $\rho_*(x)$, (\ref{l7}), equals $1$ and the convergence towards this value is clearly seen
in Fig. 5.  Time evolution of  the  pdf second moment is strongly correlated with  the fidelity interval boundary value $a$,  c.f. the left panel and compare integration results below:
\be
<X^2>_a=\int\limits_{-a}^ax^2\rho_*(x)dx=\left\{
                                           \begin{array}{ll}
                                             1, & \hbox{$a=\infty$;} \\
                                             0.95759, & \hbox{$a=30$;} \\
                                             0.93644, & \hbox{$a=20$;} \\
                                             0.87352, & \hbox{$a=10$;} \\
                                             0.75191, & \hbox{$a=5$;} \\
                                             0.45018, & \hbox{$a=2$.}
                                           \end{array}
                                         \right.\label{l9}
\ee

We note here that the fidelity interval is much broader for Cauchy
type functions than that for the  Gaussian one. This is related to
the fact that Cauchy type functions have much slower power-law decay
then Gaussians.

 The  middle panel in Fig. 5 displays the time
dependence of the pdf $\rho(x,t)$ second moment for various
$\varepsilon_2$ and the fidelity interval  $[-20,20]$.
 Like in the
gaussian  case,  the  target  pdf (\ref{l7}) is achieved faster  if
we  increase  $\varepsilon_2$.

 The  direct     and path-wise
(indirect)  methods of solution  of the master equation (3)  are
depicted in the right panel  of Fig. 3,  for three specific choices
of L\'{e}vy drivers, namely $\mu = 0.5, 1, 1.5$.     The agreement
is good and can be easily made  better if the trajectory statistics
data  are increased to more than  $200 000$ trajectories and/or the
fidelity interval gets  increased in the direct solution method.

\subsection{Cauchy target: non-Langevin modeling of Boltzmannian  equilibration.}

\begin{figure}[h]
\begin{center}
\centering
\includegraphics[width=50mm,height=50mm]{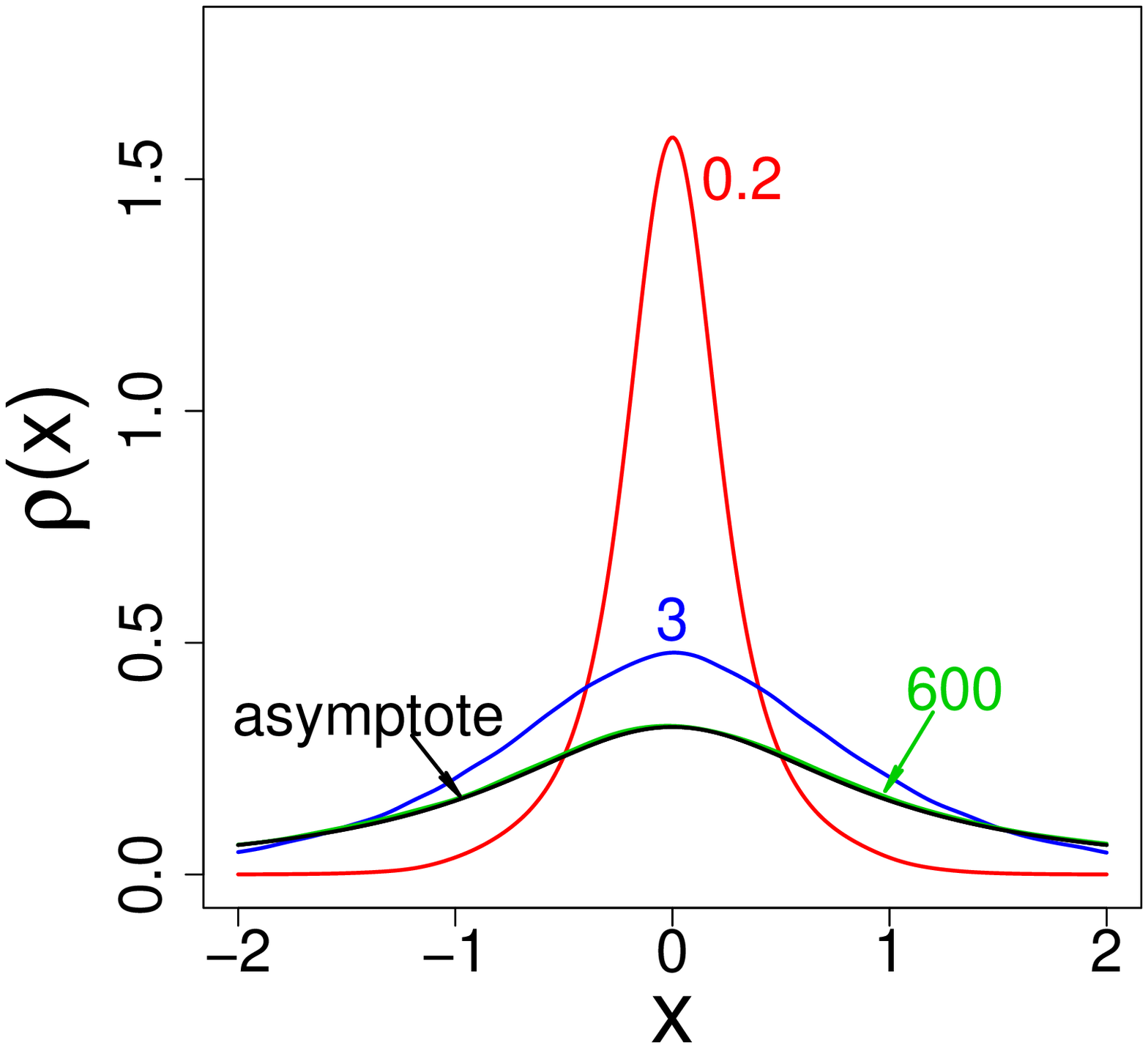}
\includegraphics[width=50mm,height=50mm]{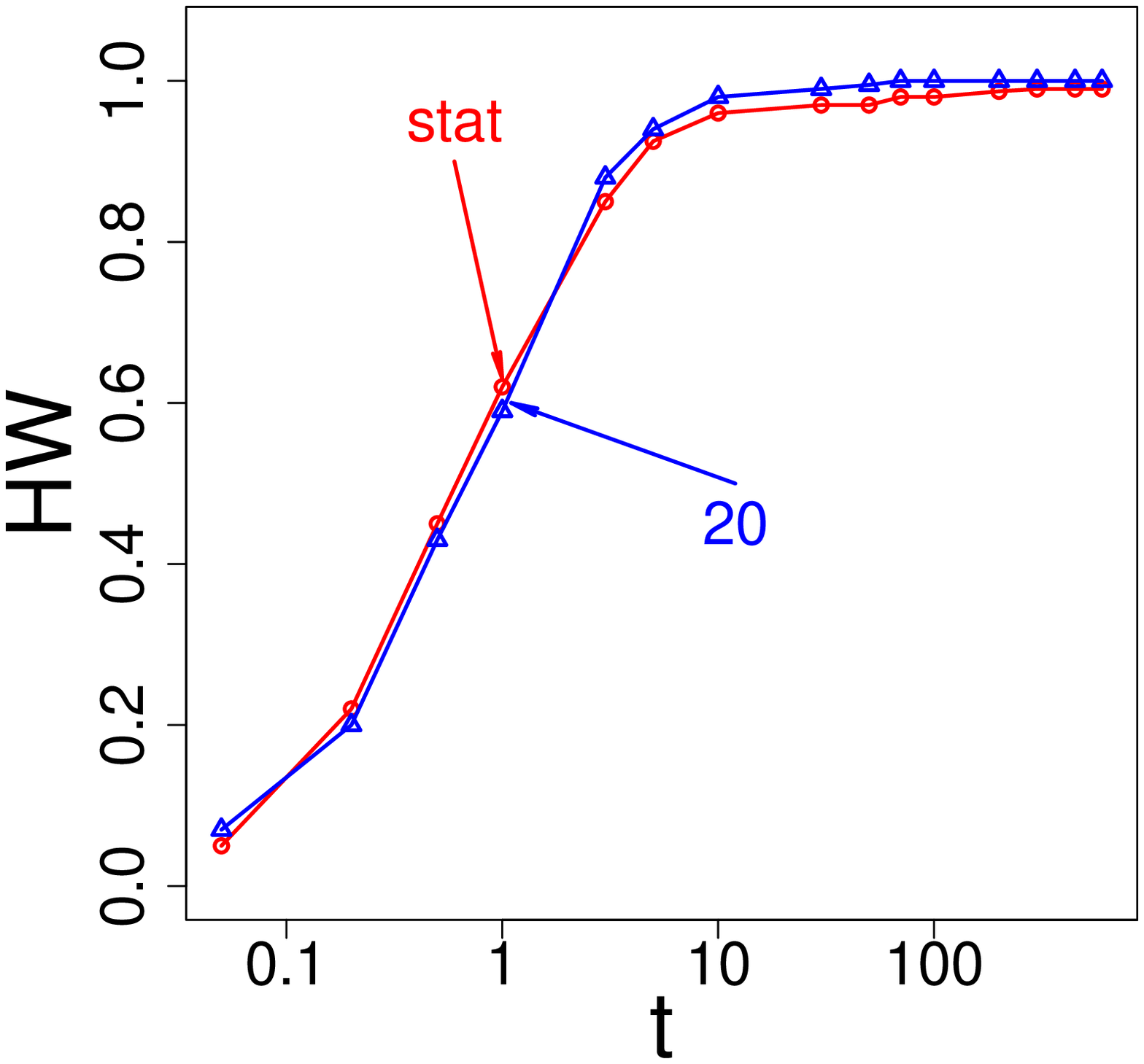}
\caption{Cauchy target. Comparison of the time evolution of $\rho(x, t)$ inferred from 200 000 trajectories
 and directly from (3). They are indistinguishable on the  left panel.  The half-width (HW)  temporal behavior is depicted in the right  panel.  The initial condition is point-wise  $\rho(x,t=0)   \sim \delta(x)$.}
\end{center}
\end{figure}

Now we consider an  asymptotic  (target) pdf to be given in  the
familiar Cauchy form. The  dynamical scenarios  that asymptotically
set down at the Cauchy pdf,  may serve as a useful diagnostic tool
when comparing the intricacies (similarities and differences)  of
the Langevin and non-Langevin equilibrating dynamics, with common
initial data and the common   terminal pdf.

 Right at this point the
major difference can be seen between the non-Langevin motion
scenario, for which the Bolztmannian (e.g. thermal) equilibration is
possible, and the standard Langevin modeling,    to which the
Eliazar-Klafter no-go statement  readily extends, see  e.g.
\cite{zaba1,klafter}.

The classic Cauchy density  is  associated both with the free Cauchy noise  (probability distribution of random jumps)  and/or the  confining \
 Ornstein-Uhlenbeck-Cauchy process  \cite{olk}. We   have
\be
\rho_*(x)=\frac{1}{\pi}\frac{1}{1+x^2}.\label{l10}
\ee
In this case, the transition rate from $x$ to $x + z$ reads
\be
w_\phi(z+x|x)=\frac{C_\mu}{|z|^{1+\mu}}\sqrt{\frac{1+x^2}{1+(z+x)^2}}.\label{l11}
\ee

We consider  the  Cauchy driver, e.g.  $\mu = 1$.   In Fig. 6 we report the   directly resolved  time evolution of   the corresponding  pdf,    its half-width (the  second moment
does not exist in the present case) for different
time instants.   The left panel  in  Fig. 6 displays results evaluated directly from (3), for the fidelity interval    $[-20,20]$.
The path-wise (indirect method) outcome  inferred from  200 000 trajectories
is indistinguishable within the scales adopted.  An approach to the asymptotic   Cauchy  pdf  is
clearly seen, together with a convergence of a half-width
to its asymptotic value $1$. The right panel of Fig. 6 compares  the    half-width of $\rho(x,t)$  temporal behavior,  obtained by means  of direct (indicated by {$20$) and path-wise ({\it stat})  methods of soplution of Eq. (3).
 Slight differences in the curves result from numerical approximations.

{\bf  Remark:}  Displayed  path-wise inferred  curves in Fig. 6 (right panel) are hampered by certain errors. The figures have been read from a histogram of
randomly sampled data.  A partitioning    into subintervals,  resulting  in the histogram shape,    is a source of inaccuracies.  One more inaccuracy source in the fine
partition case comes from the maximum read-out imprecision.  Therefore the partitioning finesse  always needs an optimization.
The time evolution of pdf,  obtained directly from the    master equation  (3),   depends on the fidelity interval adopted. This choice contributes to
 maximum and half-maximum of the curve read errors.

\subsection{Cauchy target:   Langevin-type (OUC) equilibration.}

To make a clear distinction between the  Langevin and non-Langevin
modeling, we shall describe an alternative equilibration scenario,
realizable as rather standard   Ornstein-Uhlenbeck-Cauchy  (OUC)
process, where the Bolzmann thermalization is excluded by the
argument of Ref. \cite{klafter}.

Namely, with  a  slighly modified   Cauchy density, we can associate  not only the free Cauchy noise, but also  the  confining
 OUC  process. Its
drift is given by $b(x) =   - \nabla    \Phi (x) =     - \gamma x$,
where we have adopted $\Phi $ as the Newtonian harmonic potential.
An asymptotic invariant pdf associated with  the standard fractional
(Cauchy) Fokker-Planck equation
\begin{equation}\label{zel}
\partial _t \rho = - \lambda |\nabla | \rho + \nabla [(\gamma x)\rho ]
\end{equation}
 reads:
\begin{equation} \label{couc}
\rho _*(x) = {\frac{\sigma }{\pi (\sigma ^2 + x^2)}},
\end{equation}
where $\sigma = \lambda /\gamma $, c.f. Eq. (9) in Ref. \cite{olk}.   In accordance with the Eliazar-Klafter no-go statement,  \cite{klafter},  the equilibrium pdf
does not admit a standard Boltzmannian form $\rho _*(x)   \sim \exp [-\Phi (x)]$.

 Since the OUC pdf (16)  has no variance, in Fig. 7 we visualize the temporal evolution of $\rho(x,t)$, started from
$\rho(x,t=0) \sim \delta(x)$      for two motion scenarios
corresponding to the Langevin (OUC) and  non-Langevin  cases,
provided they share a common Cauchy target pdf.
 In addition  to the direct comparison of the shapes of $\rho (x,t)$
in above evolution scenarios, for both cases we plot  the width of
the OUC "bell" at its half-maximum at a number of consecutive
instants of time. The latter can serve as an  diagnostic tool,
differentiating between those \it  definitely inequivalent \rm
motions, see also Refs. \cite{gar1,gar2}.

\begin{figure}[!h]
\begin{center}
\includegraphics [width=50mm,height=50mm]{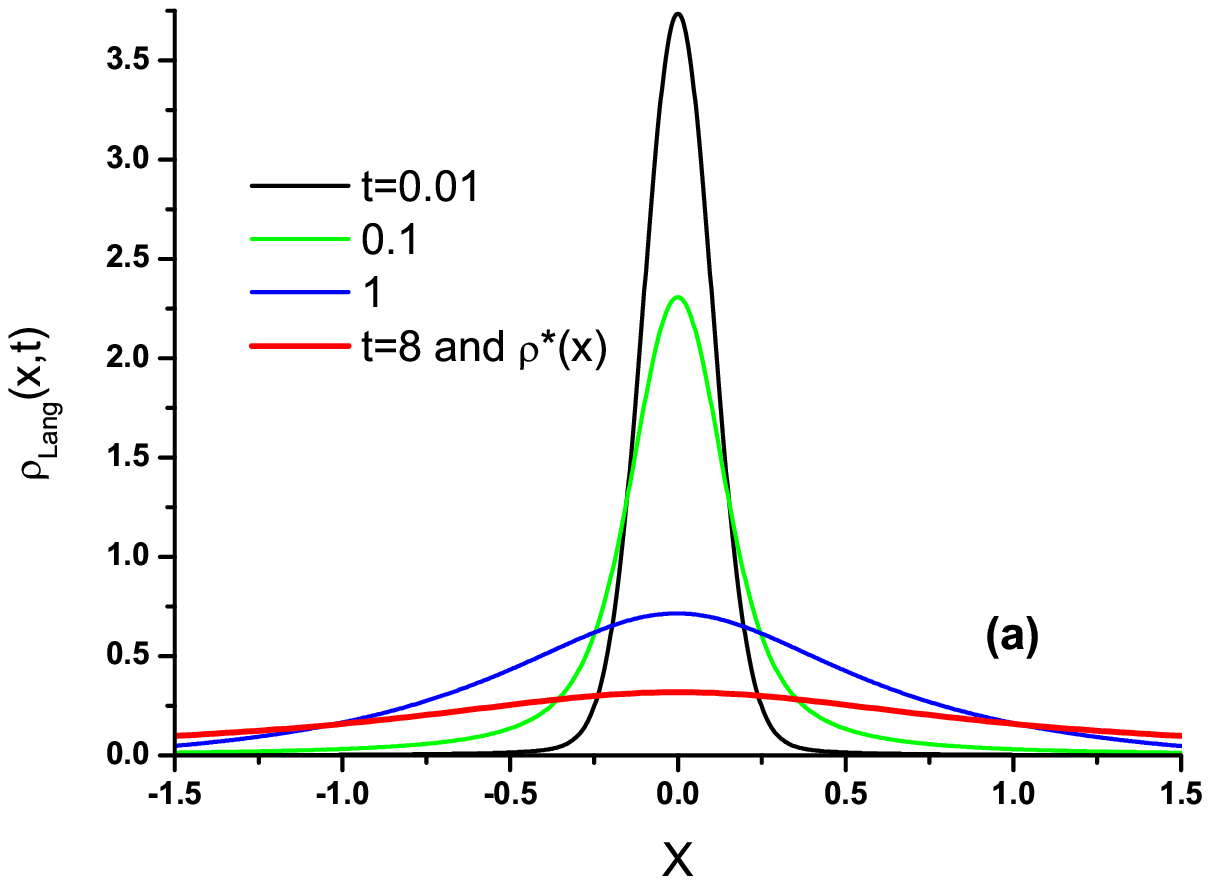}
\includegraphics [width=50mm,height=50mm]{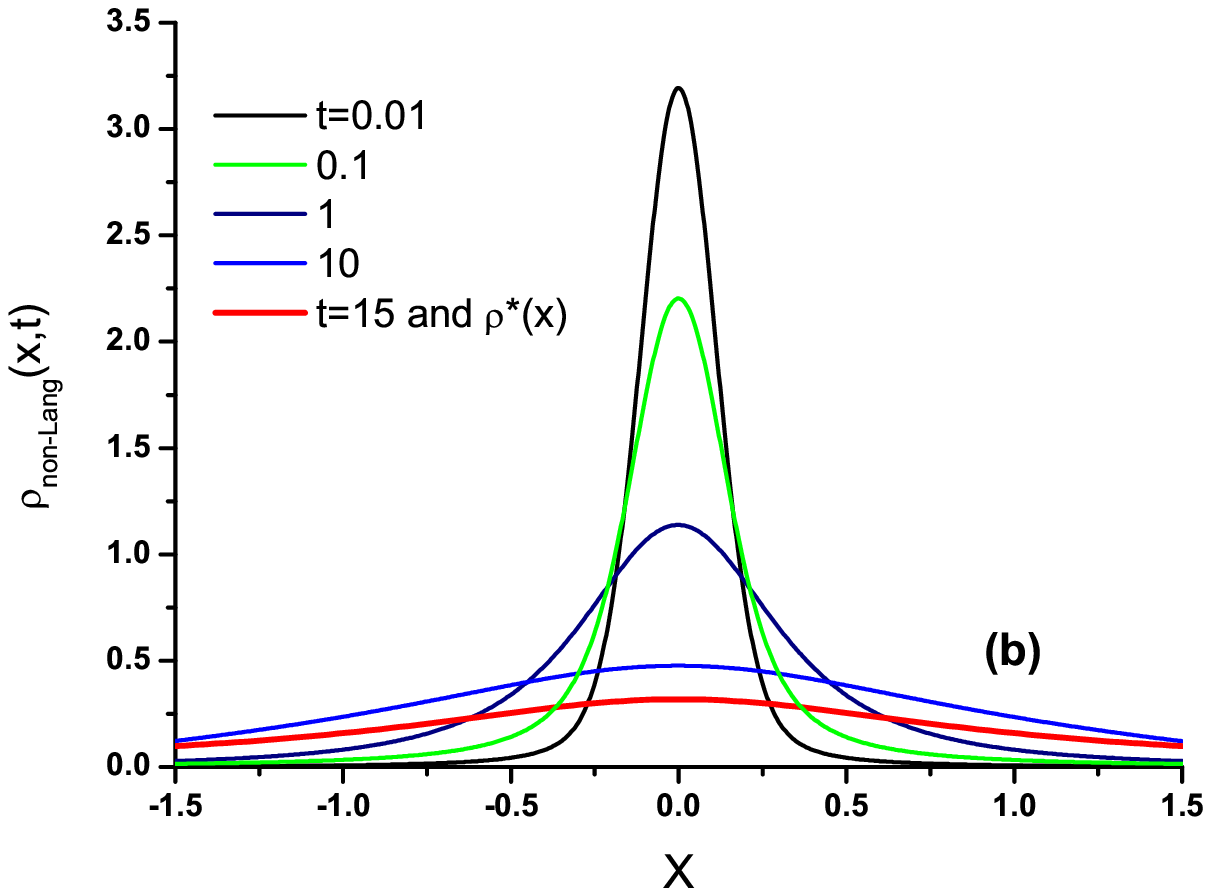}
\includegraphics [width=50mm,height=50mm]{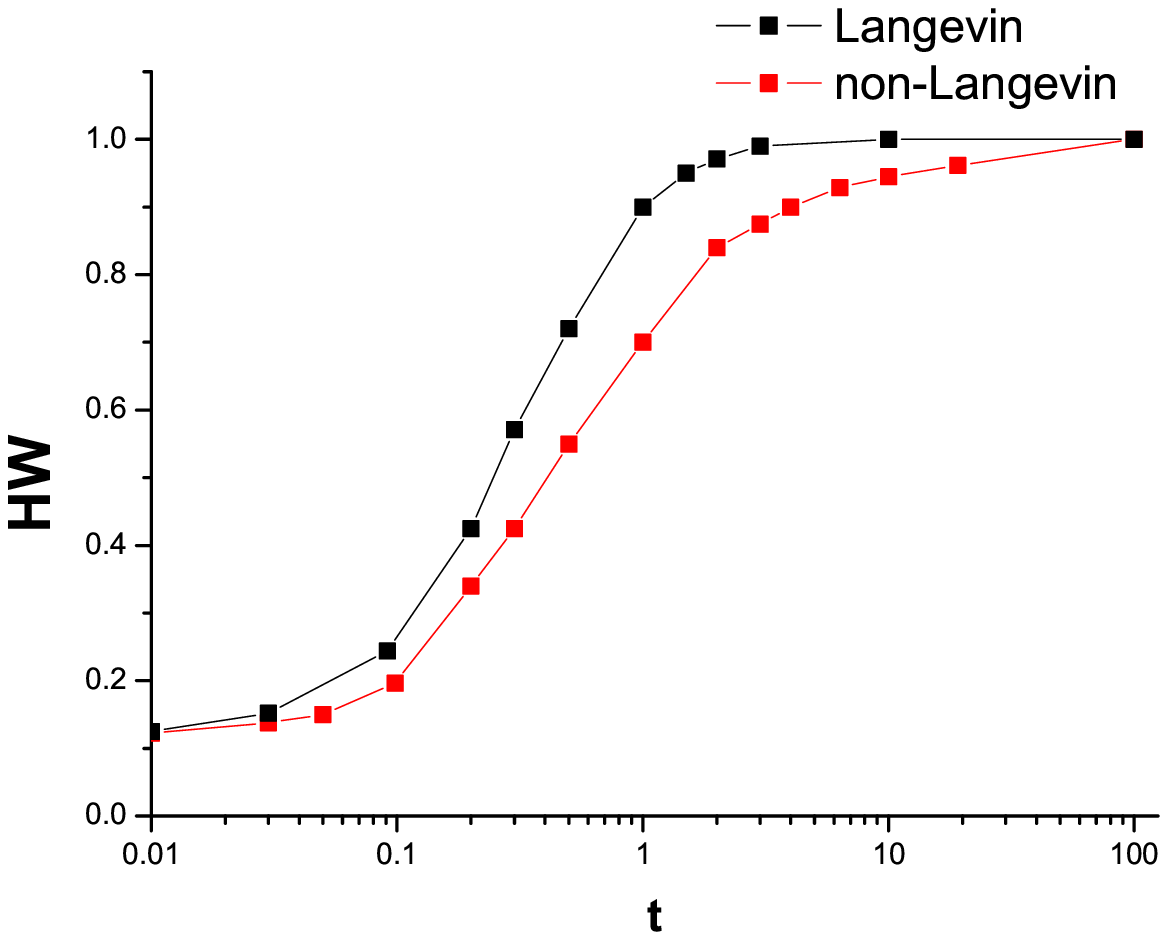}
\end{center}
\caption{Time evolution of pdf's $\rho(x,t)$ for the Cauchy-Langevin
dynamics (panel (a)). The non-Langevin  motion scenario is reported
in panel (b). For both scenarios, the common target pdf is the
Cauchy density, while the initial $t=0$ pdf is set to be  a Gaussian
with height 25 and half-width $\sim 10^{-3}$. The first depicted
stage of evolution corresponds to  $t=0.01$.  Panel (c) reports a
difference  between  the Langevin and non-Langevin patterns
 of dynamical behavior in terms of the half-width of the corresponding pdfs. }
\end{figure}

\section{Outlook}

We have completed a compatibility test for two different procedures
that can be employed to solve the master equation (1).  By the
construction we deal with  non-Langevin  jump-type processes, whose
jump intensity is induced by a properly biased L\'{e}vy - stable
noise.

In the present paper, without limiting a generality of results, we
have confined   our  attention  to the most difficult (from the
convergence point of view) case of the  non-Gaussian Cauchy driver
and its varied forms of  response to external potentials.  In
particular, for an exoplcit choice of the $\mu =1$ Cauchy driver, we
have discussed a difference between the Langevin and non-Langevin
dynamical patterns of behavior. Those,  while being started from the
same initial (Dirac delta-like) pdf data, ultimately result in the
common for both asymptotic Cauchy target pdf. Albeit following
inequivalent time evolution patterns of behavior.

  Within the non-Langevin
modeling  an emergence of   the Cauchy    target  pdf  can be
consistently interpreted   as an asymptotic outcome of the thermal
equilibration process, see e.g. also Ref \cite{thermal}. The
Boltzmann form of the pdf is fully compatible   with  the
non-Langevin framework. To the contrary, the standard Langevin
modeling is known to be incompatible with the Boltzmannian
 equilibration, c.f. \cite{klafter}.

We have emphasized before  that a leading physical motivation an
investigation of non-Langevin jump-type processes, compatible with
the master equation (1), was an observation \cite{thermal} that
L\'{e}vy type pdfs can be receive an interpretation of Boltzmann
equilibria. However, the price to be paid is that we must go beyond
the tenets of a popular   paradigm of the  Langevin  modeling of
L\'{e}vy processes.

One may argue that a trui interest of the method is its application
to stochastic problems of  non-physical origin, like econ omics or
population dynamics. A standard argument is that if those come from
physics a Langevin formulation seems to be  appropriate  and
adequate.  Our point is that the Langevin modeling  of L\'{e}vy,
quite against the standard lore, is inadequate. More than that, the
evident spatial non-locality present in the definition of noise
generators as classified by the L\'{e}vy-Khintchine formula, needs a
careful exploration as well, see e.g. \cite{jmp}.

Coming back to purely physical arguments, we note here that the
non-Langevin master equation \eqref{l1} may be interpreted as a
direct analog of the Boltzmann kinetic equation derived for pairwise
interparticle interactions \cite{land10}. This equation has been
successfully applied to describe  nonequilibrium properties of
different physical systems  like  dielectrics  or  superconductors.
It  drives   a kinetics of first order phase transitions
\cite{land10}.   Another  field, where the non-Langevin master
equation \eqref{l1} has been in fact used, is the theory of radio
spectroscopic methods, like election paramagnetic and nuclear
magnetic resonances.

The master equation of the form  \eqref{l1} is used as well  in the
description of spin-lattice relaxation in solids, liquids and gases.
These processes  originate from  spins motion, induced by an
external (probing) magnetic field. They lead to the  so-called
dynamic broadening of resonant lines, \cite{abragam}.

The very  same  master  equation,  with or  without the  detailed
balance principle imposed to hold true,   has been applied to the
problem of so-called statistical mechanics of money \cite{yak}. This
problem, related to the application of methods of statistical
physics in economy (so-called econophysics \cite{stanley}), deals
with master equation \eqref{l1} with properly (due to peculiarities
of the specific problem under consideration) modified transition
rates \eqref{l2}.

Namely, in Ref. \cite{yak}, Eq.  \eqref{l1} has been solved
numerically to obtain the outcome of the "thermal" equilibration of
the process of the random exchange of money between several agents.
The outcome turns out to be of exponential, i.e. non-Gaussian form.
The results of numerical solution of the Boltzmann-type  equation
have  been compared in Ref. \cite{yak} with direct, trajectory-wise
numerical simulations, performed earlier \cite{iskra}. The methods,
developed in the present paper, are capable to model adequately the
above class of problems.

Our present study shows clearly that the direct numerical solution
of the  master equation \eqref{l1} is sufficient to obtain the
reliable  information about the stochastic dynamics in question.
Although, as a matter of principle,   we may not  even  invoke the notion
of random trajectories underlying the  pertinent   pdf dynamics,
given the master equation data - we can always retrieve them  (e.g. trajectories)
 back.
The Gillespie's algorithm is a useful tool to this end, as
demonstrated in Ref.  \cite{zaba1}.

We emphasize that it is sometimes  useful to  resort to the indirect
 path-wise methods, based on the above mentioned algorithm.
 Specifically it is the case, when  the master equation
  of  the  type \eqref{l1}  is  a priori  known  and one attempts
  to  devise (model)   or   deduce    an underlying    microscopic model  of  random motion.  This
  viewpoint is particularly important if   the  standard  Langevin modeling
  does    fail.

To compare the efficiency   and computer time consumption   of both
direct and indirect algorithms to solve (1) is not trivial at all
and cannot be settled unambiguously.

A direct  integration of Eq. (1) needs some care regarding the
spatial interval $[-a,a]$  and time  interval (hence fixing a
sufficiently small value of $\Delta t$) partitioning finesse. The
ultimate computer time  consumption  depends quite sensitively on
the proper balance  between those two partitioning options.

The path-wise method
 appears to be  much more computer time  demanding  than  the  direct kinetic modeling.
A  dynamical  retrieval  of   a single  $\rho (x,t)$,   needs so
many separate runs  in  the      path-wise  approach,  that the
computer simulation and subsequent  acquisition of  sufficiently
large statistical data   may  take  not merely minutes,
      but days or weeks.
Typically  the  path-wise method needs a simulation and accumulation
(data storage) of a large number of trajectories, typically between
$10^3$ and $ 3 \cdot 10^3$. Even with the parallel computing on
small clusters the procedure may take days, for larger values of the
stability index $\mu \in (0,2)$. Especially when increasing the
spatial dimensionality from $1d$ to $2D$, \cite{zaba2},  or $3D$.

The major gain of the path-wise method is that we can  adequately
simulate a microscopic motion scenario that is compatible with the
dynamics (1).

  {\bf Remark:}  Since quite varied (here Langevin and non-Langevin  respectively) microscopic
    motion scenarios can  be used to  implement a dynamical
    interpolation between any  given a priori pair of initial (here
    narrow Gaussian) and target   pdf data, we get  a  reason   to
    mention and
    advocate a complementary viewpoint,  based on the exploitation
    of diffusion-type processes, instead of jump-type ones.
    In this connection, we mention our recent \cite{gar2}
    finding that, under suitable confinement conditions
    (i.e. properly tailored external potentials) and  in clash with the folk
    lore ways of thinking in this context,   the  ordinary
     Fokker - Planck equation may generate non-Gaussian heavy-tailed pdfs,
      (like Cauchy or more general L\'{e}vy-stable distributions) in its long-time
       asymptotics.  Specifically, the  problem of physical relevance  is,  whether the three dimensional
       Fokker - Planck equation,  in the presence of a specific external potential,
          can be employed  to describe the non-ergodic  long-time relaxation in glassy systems. This issue
            is   presently  under investigation.

\end{document}